\documentstyle[prl,aps,twocolumn,epsf]{revtex}

\begin{document}
\pagestyle{empty}

\noindent
{\bf Comment on ``Two Time Scales and Violation of the
Fluctuation-Dissipation Theorem in a 
Finite Dimensional Model for Structural Glasses''}

\begin{figure}
\begin{center}
\mbox{\epsfysize=5.5cm\epsfbox[50 40 517 507]{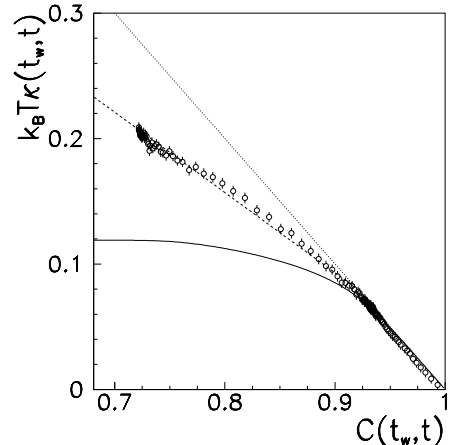}}
\end{center}
\caption{Circles: response versus correlation for size
$32^3$, $\beta J=\infty$ and $\beta\mu=10$.
Dashed straight line: same fit of Ref.~\protect\cite{jeff}.
Solid line: function corresponding to the equilibrium $P(q)$ plotted
in Fig. \ref{fig2}. The dotted line is $1-x$.}
\label{fig1}
\end{figure}

In a recent paper, Ricci-Tersenghi {\em et al.} \cite{jeff}
show that, in the frustrated Ising lattice gas (FILG) \cite{mimmo},
the fluctua\-tion-dissipation relation between density-density
correlations and associated responses, following a quench in the
chemical potential, is characterized by two
linear regimes, with a piecewise constant fluctuation-dissipation ratio (FDR).
The authors take the value $\beta J=10$ of the coupling between particles,
and quench the system from a low value of $\beta\mu$, in the liquid phase,
to $\beta\mu=10$, deep in the glassy phase.
After a waiting time $t_w$ they apply a
small random perturbation in the chemical potential,
and measure the correlation $C(t_w,t)$ and the integrated
response $T\kappa(t_w,t)$.
For $t-t_w\ll t_w$ they find the linear FDT regime
$T\kappa(t_w,t)=1-C(t_w,t)$, and for $t-t_w\ge t_w$
the linear out of equilibrium regime
$T\kappa(t_w,t)=(1-q_{EA})+X(q_{EA}-C(t_w,t))$,
with $X=0.64(3)$ and $q_{EA}=0.92(1)$.
We have repeated the experiment for size $32^3$, 
taking this time for convenience the value $\beta J=\infty$
and quenching to the
same value $\beta\mu=10$, with $t_w=10^5$ and a perturbation
$\beta\epsilon=0.1$.
The result is shown in Fig.~\ref{fig1} (circles):
it is compatible within the errors with the two linear regimes
found in Ref.~\cite{jeff} (dashed line).
Here we want to suggest that it is plausible that this behavior does not
correspond to the asymptotic regime, which should be given by the
solid line of Fig.~\ref{fig1}.

\begin{figure}
\begin{center}
\mbox{\epsfysize=5.5cm\epsfbox[50 40 517 507]{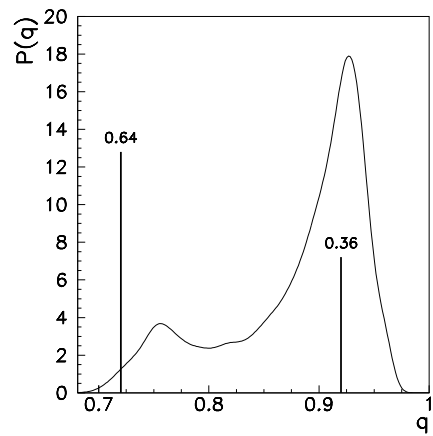}}
\end{center}
\caption{Solid line: equilibrium distribution $P(q)$
for size $10^3$, $\beta J=\infty$, $\beta\mu=10$.
Vertical bars: function
$X\delta(q-q_{\text{min}})+\mbox{}$ $(1-X)\delta(q-q_{EA})$
with the same $X$ and $q_{EA}$ of Ref. \protect\cite{jeff}.}
\label{fig2}
\end{figure}

It has been recently proved \cite{peliti} that,
for systems in which
the free energy density tends asymptotically to the
equilibrium one, and in which the stochastic stability holds,
the FDR function $X(q)$ is connected to the 
equilibrium overlap distribution $P(q)$. The piecewise constant FDR
found in Ref. \cite{jeff} would then correspond,
for the density overlap of the FILG,
to the distribution
$P(q)=X\delta(q-q_{\text{min}})+(1-X)\delta(q-q_{EA})$,
where $q_{\text{min}}=C(t_w,t\to\infty)$.
We have measured $P(q)$ on a system of size $10^3$ using the
Parallel Tempering technique, with $\beta J=\infty$ and 25 chemical potentials
between $\beta\mu=1$ and $\beta\mu=10$, averaging over 32 disorder
configurations. We have checked the thermalization of the system looking
at the symmetry of the spin overlap distribution.
In Fig. \ref{fig2} we show the result for $\beta\mu=10$ (solid line),
together with the two
delta functions corresponding to the FDR of Ref. \cite{jeff}.
Of course
for finite size the delta functions will be smeared out, and one must expect
two non-zero width peaks, but also taking this in account the two
distributions are definitely different.
For a crossed check, 
in Fig. \ref{fig1} we plot the function that would correspond
to the equilibrium $P(q)$ (solid line):
it is definitely not compatible with the measured $T\kappa(t_w,t)$
versus $C(t_w,t)$.

It is important to point out that the overlap distribution 
$P(q)$ shown here has been measured using a particularly efficient algorithm,
and taking care that the system had indeed thermalized.
Therefore we believe that the susceptibility deduced from the $P(q)$
is more reliable than the one measured in the off-equilibrium experiment.
Indeed the latter may suffer from a number of problems, for example
that at $t_w$ one time observables have not yet reached the equilibrium
values.

\noindent 
A. de Candia and A. Coniglio,\\
Universit\`a di Napoli ``Federico II''\\
INFM, Unit\`a di Napoli

\end{document}